\documentclass[5p,twocolumn]{elsarticle}

\usepackage{amssymb}

\journal{Trans. JSASS Aerospace Tech. Japan}

\begin{document}

\begin{frontmatter}



\title{Design of a silica-aerogel-based cosmic dust collector for the Tanpopo mission aboard the International Space Station}


\author[First,Second]{Makoto Tabata\corref{cor1}}
\ead{makoto@hepburn.s.chiba-u.ac.jp}
\cortext[cor1]{Corresponding author.} 
\author[Third]{Eiichi Imai}
\author[First]{Hajime Yano}
\author[First]{Hirofumi Hashimoto}
\author[Second]{Hideyuki Kawai}
\author[Fourth]{\\Yuko Kawaguchi}
\author[Fifth]{Kensei Kobayashi}
\author[Sixth]{Hajime Mita}
\author[Seventh]{Kyoko Okudaira}
\author[Eighth]{Satoshi Sasaki}
\author[Ninth]{\\Hikaru Yabuta}
\author[Fourth]{Shin-ichi Yokobori}
\author[Fourth]{Akihiko Yamagishi}

\address[First]{Institute of Space and Astronautical Science (ISAS), Japan Aerospace Exploration Agency (JAXA), Sagamihara, Japan}
\address[Second]{Department of Physics, Chiba University, Chiba, Japan}
\address[Third]{Department of Bioengineering, Nagaoka University of Technology, Nagaoka, Japan}
\address[Fourth]{Department of Applied Molecular Biology, Tokyo University of Pharmacy and Life Sciences, Hachioji, Japan}
\address[Fifth]{Department of Chemistry, Chemical Engineering and Life Science, Yokohama National University, Yokohama, Japan}
\address[Sixth]{Department of Life, Environment and Materials Science, Fukuoka Institute of Technology, Fukuoka, Japan}
\address[Seventh]{Office for Planning and Management, University of Aizu, Aizu-Wakamatsu, Japan}
\address[Eighth]{School of Bioscience and Biotechnology, Tokyo University of Technology, Hachioji, Japan}
\address[Ninth]{Department of Earth and Space Science, Osaka University, Toyonaka, Japan}

\begin{abstract}
We are developing a silica-aerogel-based cosmic dust collector for use in the Tanpopo experiment to be conducted on the International Space Station. The mass production of simple two-layer hydrophobic aerogels was undertaken in a contamination-controlled environment, yielding more than 100 undamaged products. The collector, comprising an aerogel tile and holder panel, was designed to resist launch vibration and to conform to an exposure attachment. To this end, a box-framing aerogel with inner and outer densities of 0.01 and 0.03 g/cm$^3$, respectively, was fabricated. The aerogel mounted in the panel passed random vibration tests at the levels of the acceptance and qualification tests for launch. It also withstood the pressure changes expected in the airlock on the International Space Station.
\end{abstract}


\end{frontmatter}


\section{Introduction}
\label{}
The Tanpopo (meaning ``dandelion'' in Japanese) mission is an astrobiological experiment to be conducted on the Japanese Experiment Module (JEM) on the International Space Station (ISS) \cite{cite1}, \cite{cite2}. The primary goal of the mission is to determine whether terrestrial microbes in dusts ejected by events such as volcanic eruptions can reach the low Earth orbit (LEO) and whether interplanetary dust particles (IDP) containing prebiotic organic compounds can migrate among solar system objects. To achieve this goal, we plan composite experiments using the Exposed Facility (EF) in the JEM. These experiments will sample cosmic dust returning from and assess microbe and organic compounds that are exposed in the LEO. After a one-year sampling and exposure period, the returned samples will be biochemically analyzed in our ground laboratories \cite{cite3}, \cite{cite4}, \cite{cite5}. The density of microbes in the LEO (if found) will be determined, and intact IDPs that have not been contaminated and damaged by the Earth's atmosphere will be analyzed.

We have developed an ultralow-density (0.01 g/cm$^3$) silica aerogel for capturing hypervelocity microparticles \cite{cite6}, which realizes an almost intact collection of cosmic dusts. Cosmic dusts, including terrestrial dusts, IDPs, and space debris, impact the ISS capture medium at velocities of up to $\sim $10 km/s. Silica aerogel, a colloidal form of quartz (SiO$_2$), is an eminently suitable medium for cosmic dust sampling because of its light weight and optical transparency \cite{cite7}, \cite{cite8}. Our aerogel was especially designed for the Tanpopo mission and is not contaminated by bacterial deoxyribonucleic acids (DNA) \cite{cite9}. The material is formed as an integrated monolithic tile with different density layers and box-framing structures \cite{cite10}. Being hydrophobic, it also suppresses age-related degradation caused by moisture absorption \cite{cite11}.

Because the aerogel-based cosmic dust collector will be mounted on the Exposed Experiment Handrail Attachment Mechanism (ExHAM), newly developed by the Japan Aerospace Exploration Agency, and equipped to the EF using a robotic arm and airlock on the JEM, it requires a special aerogel holder. The aerogel loaded in the holder must pass a vibration test for launch and a pressure test. We plan to expose multiple aerogels to several different sides of the ISS, both parallel and perpendicular to its direction of travel (i.e., the East, which is the direction of the ISS orbit, West, North, and zenith direction) in three-time annual exposures. In this paper, we present the conceptual design and development of the aerogel cosmic dust collector for the Tanpopo mission.

\section{Mass-production Tests of Aerogels}
\label{}
We conducted a series of test aerogel productions at Chiba University from April to November in 2011. The test production provided a feasibility study of mass production by our current technique. Ideally, multilayer aerogels should be systematically produced at high yield to meet the demand for multiple exposures of the aerogels. Thus, we plan to deliver a maximum of 48 aerogel tiles to the ISS. Furthermore, they should be carefully manufactured in a contamination-controlled environment. The production test yielded over 100 undamaged aerogel products.

\subsection{Production method}
Aerogels were manufactured by the KEK method \cite{cite11}. By this method, we can produce hydrophobic aerogels of densities between 0.02 and 0.5 g/cm$^3$. The key procedures of the aerogel production are wet-gel synthesis, hydrophobic treatment, and supercritical drying. First, a wet gel is synthesized by hydrolyzing, condensing, and polymerizing polymethoxysiloxane in a mold. An ammonia aqueous solution is used as the catalyst and ethanol is used as the solvent. After aging, the wet gel is immersed in ethanol and detached from the mold. Next, the gel is rendered hydrophobic and is washed three times in ethanol, replacing the ethanol at each wash. Finally, the wet gel is supercritically dried using supercritical extraction equipment, including an autoclave. The production of a single batch takes 25 days. The full procedure of aerogel production is described in Tabata \textit{et al.} (2012) \cite{cite11} and our technique, focusing on ultralow-density aerogels, is described in Tabata \textit{et al.} (2011) \cite{cite9}.

Multilayer aerogels are formed at the wet-gel synthesis stage. Once the first wet-gel layer is gelled, the second layer can be deposited on the first layer without mixing. The fabrication of multilayer aerogels has been described in Tabata \textit{et al.} (2005) \cite{cite10} and Tabata \textit{et al.} (2011) \cite{cite9}. The mold is a polystyrene case, from which the wet gel is readily detached for subsequent hydrophobic treatment and supercritical drying.

The production process involves the transformation of the originally hygroscopic aerogels into hydrophobic materials (hydrophobic treatment for wet gels). A successfully produced hydrophobic ultra-light aerogel as an independent monolithic tile of density 0.01 g/cm$^3$ is reported in Tabata \textit{et al.} (2010) \cite{cite6}.

The wet gels are supercritically dried in carbon dioxide (for details, see Tabata \textit{et al.} (2012) \cite{cite11}). For low-density aerogels, carbon dioxide can be replaced with ethanol, as implemented in Tabata \textit{et al.} (2011) \cite{cite9}. This is an advantage because our supercritical extraction equipment for ethanol is easier to operate than our equivalent carbon dioxide apparatus.

\subsection{Aerogel specification}
Commercially available square cases (sides 96 mm) were used as the mold in wet-gel synthesis. At this stage, a holder for aerogels in the LEO had not been fully designed. However, as a rough estimate, we planned dimensions of 100 mm $\times $ 100 mm and a thickness of 20 mm for the aerogel.

As a product specification, we chose a simple two-layer configuration with different densities. More specifically, the configuration consists of a base layer of density 0.03 g/cm$^3$ and a top layer of density 0.01 g/cm$^3$. The total aerogel thickness is 20.5 mm, and thicknesses of the base and top layers are 10 and 10.5 mm, respectively. The additional 0.5 mm in the top layer allows for fixation to the holder, which compresses the layer. To allow for shrinkage in the production process, the wet gels were synthesized at a thickness of 21.5 mm.

The fabrication of chemically combined layer aerogels without separation into two independent tiles remains a major challenge. In general, the shrinkage of the wet gel during production depends on the aerogel density; the lower the aerogel density, the greater the shrinkage. This results in layer separation for larger-area aerogels. However, for 96-mm molds, we considered that layer separation would not present a serious problem.

\subsection{Contamination control}
Most of the aerogel production processes were performed in a clean booth (class 1000) to avoid possible contamination by dust particles, bacterial DNA, and amino acids \cite{cite9}. Washable tools were treated by the following procedure before being placed in the clean booth. First, the tools were soaked in a detergent using a manual washing kit (5\% Extran MA01, Merck) for 24 h and then well rinsed with tap water followed by ultrapure water. Finally, they were air dried in the clean booth. A laboratory bench in the clean booth was swabbed by TechniCloth clean room wipers (VWR International LLC) absorbed with 5\% Extran.

HPLC-grade ethanol (99.5\%, Kanto Chemical Co., Inc.) was introduced to minimize contamination by amino acids during production. It was used in both wet-gel synthesis and washing processes. In our previous studies, special-grade ethanol (99.5\%, Wako Pure Chemical Industries, Ltd.) and ethanol (99\%, Japan Alcohol Trading Co., Ltd.) were used in wet-gel synthesis and washing, respectively.

Carbon dioxide was employed for supercritical drying in this test production, although ethanol has been used in our previous studies. Prior to testing, we examined contamination by amino acids in aerogels dried by carbon dioxide and ethanol in a trial production. Supercritical drying by carbon dioxide yielded more pure aerogels than drying by ethanol \cite{cite12}. To prevent contaminating particles from flowing to the autoclave from the liquid carbon dioxide supply cylinder, an inline filter for liquid (Swagelok Co.) was installed in front of an input valve of the carbon dioxide supply line.

The manufactured aerogels were preserved in air-tight buckle containers. The aerogel was first stored in a washed polystyrene square case of sides 116 mm. This case was doubly packed in zippered pouches. Finally, the pouch was deposited in an air-tight container wiped by a TechniCloth soaked with Extran detergent.

\subsection{Production timeline}
We originally planned the production of 126 aerogel tiles in 14 batches. Each batch includes 9 tiles that can be obtained in a single supercritical drying operation. Because the ultralow-density aerogels are brittle and fragile, they may be damaged and cracked in the production process. Assuming that two out of every three aerogel products are undamaged, we aimed to obtain 84 pristine tiles. The start of test production was scheduled for April 2011, followed by two supercritical drying operations per month from May to November. Wet gels in two batches were synthesized over a continuous two-day period and were simultaneously treated prior to supercritical drying, as described below.

\subsection{Production results}
The layer separation problem arising in the first to fourth batches was resolved in the fifth batch by first synthesizing the top layer, aiming at a density of 0.01 g/cm$^3$. In the first four batches, the base layer was synthesized at 0.03 g/cm$^3$, followed by the top layer at 0.01 g/cm$^3$. During aging, layer separation occurred in all wet gels in the first and second batches; hence, the wet gel could not be detached from the mold. In the third and fourth batches, layer separation occurred in 11 out of 18 tiles, and only four products were successfully fabricated after supercritical drying. The 0.01 g/cm$^3$ wet gel layer failed to shrink, and therefore, it could not be separated from the mold, whereas the 0.03 g/cm$^3$ layer shrank well. The converse usually occurs in single-layer wet gels. The layer separation problem was empirically solved by inverting the top and base layers in the wet-gel synthesis process.

The failure rate of supercritical drying was reduced by replacing the pump and needle valve. The pump conveys high-pressure liquid carbon dioxide to the autoclave. The needle valve is a check valve mounted between the pump and autoclave. Supercritical drying failed in the fifth and sixth batches, resulting in shrunken and deformed aerogels. The primary reason for the failure was age-related degradation of the pump and needle valve, causing the required high pressure (beyond the critical pressure of carbon dioxide; 7.4 MPa) to decrease. To prevent similar failures, we ensure that an alternative pump is always available.

\begin{figure}[ht] 
\centering 
\includegraphics[width=0.48\textwidth,keepaspectratio]{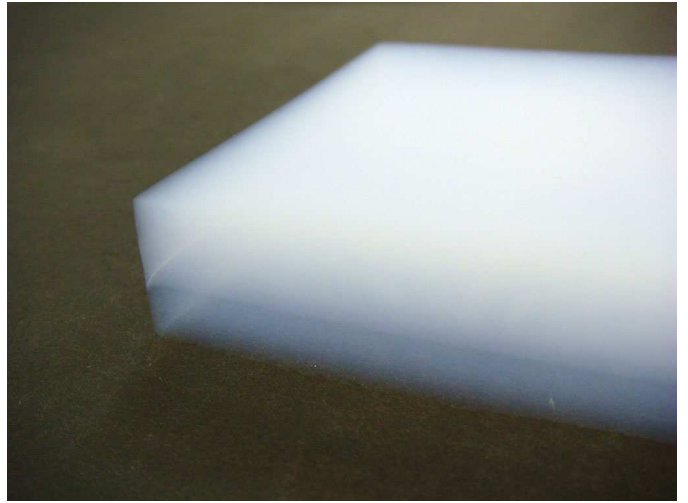}
\caption{Two-layer aerogel produced in the mass-production test (ID = TNPP p8-3a). The densities of the blue base and white top layers are 0.03 and 0.01 g/cm$^3$, respectively.}
\label{fig:fig1}
\end{figure}

\begin{figure}[ht] 
\centering 
\includegraphics[width=0.50\textwidth,keepaspectratio]{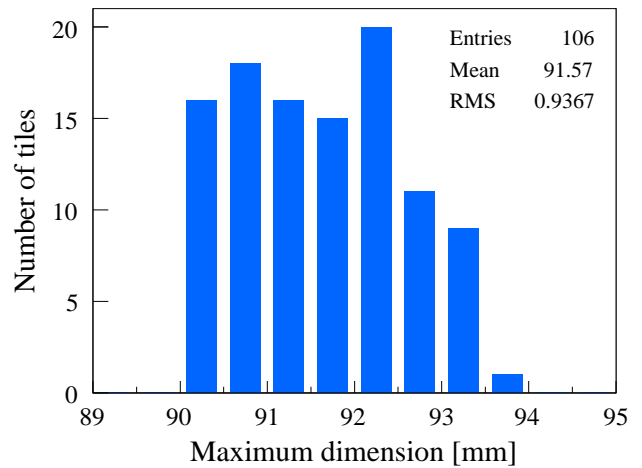}
\caption{Maximum dimension distribution of 106 undamaged aerogel tiles.}
\label{fig:fig2}
\end{figure}

The first successful aerogel tile with the modified layer configuration was obtained in the seventh batch in July 2011. Since then, aerogel products have been well fabricated up to the planned fourteenth batch. Successful fabrication was achieved in additional five batches, except for the eighteenth batch (supercritical drying failure).

A total of 106 tiles were obtained as usable products (see Fig. \ref{fig:fig1}). The total number of tiles produced after the seventh batch was 126; therefore, the yield of usable products was 84\%. Supercritical drying failed in a single subsequent batch. The average and standard deviation of the maximum dimensions of the obtained aerogels were 91.6 and 0.9 mm, respectively (see Fig. \ref{fig:fig2}), indicating a longitudinal shrinkage ratio of 95\%. The average measured thickness was 20.1 mm. Aerogel density was indirectly obtained by measuring the refractive indices of a sample tile (ID = TNPP p8-1b) at the tile corners using a laser operating at 405 nm (Fraunhofer method \cite{cite11}). Refractive indices of the top and base layers were 1.0029 and 1.0088, respectively. Using the conversion described in Tabata \textit{et al.} (2012) \cite{cite11}, the corresponding densities are 0.0116 and 0.0351 g/cm$^3$, respectively. The manufactured aerogels will be utilized for the following purposes: hypervelocity impact simulation experiments using a two-stage light-gas gun, contamination control study, detailed design of the aerogel holder, and establishment of analytical procedures of captured particles.

\begin{figure}[t] 
\centering 
\includegraphics[width=0.48\textwidth,keepaspectratio]{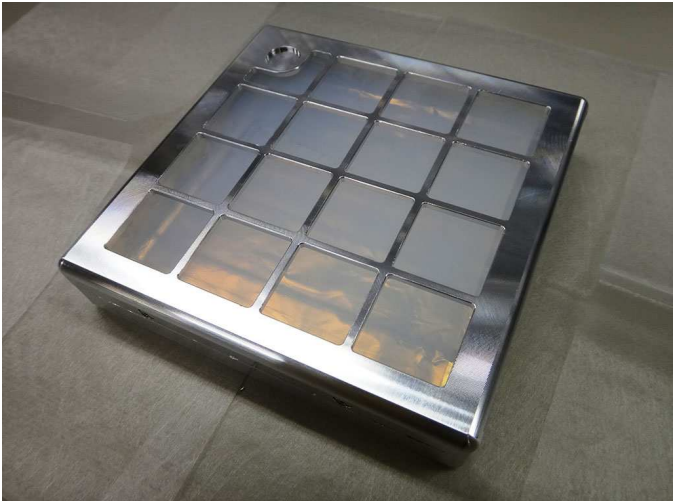}
\caption{Prototype of the sample aerogel panel, in which a prototype aerogel is mounted. The witness plate is located at the upper left. The cover plate is not shown.}
\label{fig:fig3}
\end{figure}

\section{Sample Aerogel Panel Design}
\label{}
For interfacing with the ExHAM, the outer dimensions of the aerogel holder were determined as 100 mm $\times $ 100 mm $\times $ 20 mm. We refer to the aerogel holder as the ``sample aerogel panel.'' The panel and aerogel configuration were designed, and prototypes were manufactured at the end of 2012.

\subsection{Holder design}
The sample aerogel panel was designed for maximum exposure area and aerogel thickness. Fig. \ref{fig:fig3} shows the prototype of the aerogel panel manufactured by Yuki Precision Co., Ltd. The panel, fabricated from aluminum alloy A7075, consists of a body case and gridded lid. To protect the exposed surface of aerogels, a cover plate was attached to the aerogel panel.

\begin{figure}[t] 
\centering 
\includegraphics[width=0.48\textwidth,keepaspectratio]{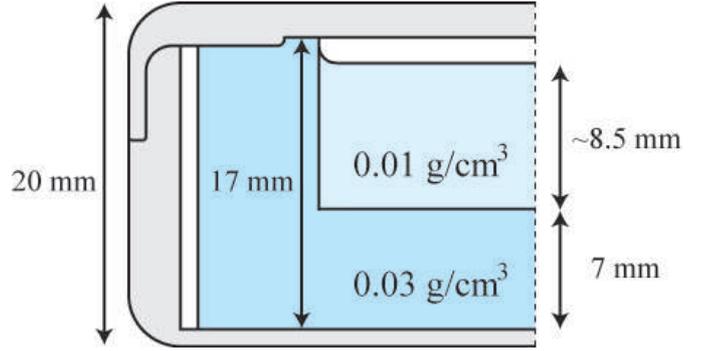}
\caption{Schematic of the edge of the sample aerogel panel and box-framing aerogel. The top and base aerogel layers (of densities 0.01 and 0.03 g/cm$^3$) are shown in pale blue and light blue, respectively. The sample panel comprising the body and lid is shown in gray. The grids are offset from the 0.01 g/cm$^3$ aerogel surface.}
\label{fig:fig4}
\end{figure}

To ensure scientific outputs and safety of the ISS crews, the grids on the lid of the aerogel panel were designed to accommodate 16 windows. The grids retain the aerogel within the panel if the aerogel disintegrates by vibrations during rocket launching or spacecraft landing shocks. Moreover, the grids minimize accidental damage to the aerogel by the edge of the panel cover or from the fingers of the ISS crews during the mounting of the cover to the panel. The aerogel sample can be analyzed provided that the aerogel remains on the panel. The grids also assist in preventing aerogel chips from being scattered aboard the ISS. To ensure adequate strength, the cross section of the grids was set at 2 mm $\times $ 2 mm. To protect the 0.01 g/cm$^3$ aerogel from potential grid vibrations, the grids must never touch the 0.01 g/cm$^3$ aerogel surface, as shown in Fig. \ref{fig:fig4}.

The total exposure area, excluding a witness plate, is approximately 56 cm$^2$ per panel. The witness plate is positioned at the corner of the farthest window to create an aerogel surface that is not directly exposed to space. Large-area windows are required for sampling as many cosmic dusts as possible during a limited time frame. Allowing a margin for tapping between the panel body and lid, the inner secured width is 94 mm. To maintain a clearance of 1 mm at each side of the aerogel, the required aerogel dimensions are 92 mm $\times $ 92 mm. The aerogel is fixed in the panel by compressing the edges (width 5 mm) with the lid (see Fig. \ref{fig:fig4}).

The maximum thickness of the aerogel was designed to be 17 mm. Thicker aerogels are required to undoubtedly capture cosmic dusts striking with a higher impact energy. The thicknesses of the bottom of the body case and the top of the lid were reduced to 1 and 2.5 mm, respectively. Specifically, the aerogel is fixed by 0.5 mm compression.

\subsection{Aerogel configuration}
A box-framing aerogel configuration was adopted in the sample aerogel panel. This configuration constitutes a top layer of density 0.01 g/cm$^3$ integrated in a 0.03 g/cm$^3$ hollow box-shaped base layer (see Fig. \ref{fig:fig4}). Unlike the simple two-layer structure, the bottom layer forms a three-dimensional frame around the top layer, surrounding its four sides as well as the base layer. The box-framing aerogel was conceptualized by Tabata \textit{et al.} (2005) \cite{cite10}. The density of the top layer (0.01 g/cm$^3$) is the lowest that our technique can produce, and it is expected to capture even volatile particles. The 0.03 g/cm$^3$ base layer should capture high-energy particles with a large diameter or high impact velocity.

The box-framing structure enhances the strength of this aerogel configuration, because the exposed top layer, which has low density and is thus easily damaged, is protected by the higher-density, more robust base layer. The high performance of aerogels with a density of 0.03 g/cm$^3$ has been previously demonstrated by Japan's LEO mission \cite{cite13}. To avoid grid-contact damage to the fragile top layer, the box-framing aerogel is fixed in the sample aerogel panel by compressing only the base; the simple two-layer aerogel was seriously damaged in preliminary vibration tests conducted in the fall of 2012.

\begin{figure}[t] 
\centering 
\includegraphics[width=0.48\textwidth,keepaspectratio]{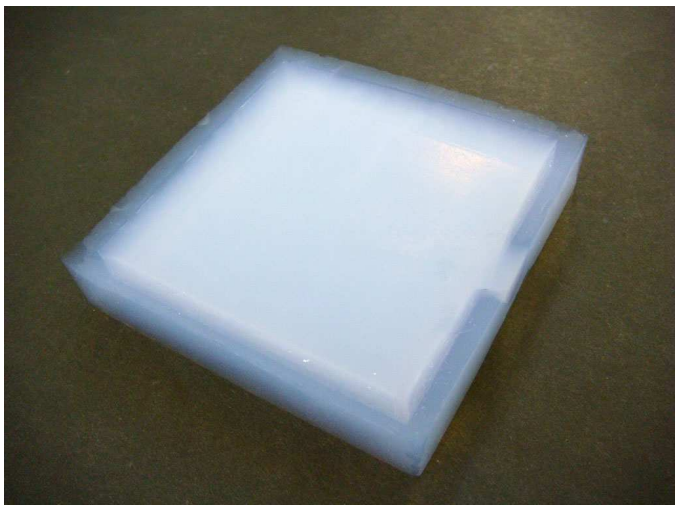}
\caption{Prototype of the box-framing aerogel, consisting of a 0.03 g/cm$^3$ base frame and a 0.01 g/cm$^3$ top layer (ID = TNPP d25-1a).}
\label{fig:fig5}
\end{figure}

A prototype of the box-faming aerogel is shown in Fig. \ref{fig:fig5}. The 0.03 g/cm$^3$ base layer was cast using a handmade plastic attachment in addition to the commercially available square case. The base was then filled with the 0.01 g/cm$^3$ top layer. The maximum width and average thickness of the base layer are 92.5 mm and 17.5 mm, respectively. Observed from the top, the average width of the side frame to be fixed in the panel is 6.5 mm. The thicknesses of the space-exposed base and top layers at the tile center are estimated at approximately 7 and 10 mm, respectively. The weight of the aerogel is 3.1 g.

\section{Vibration and Pressure Tests}
\label{}
To ensure scientific outputs and safety aboard the ISS, the prototype panel including the aerogel must pass both vibration and pressure tests before the flight models are begun. These tests were successfully performed from the end of 2012 to the spring of 2013.

\subsection{Vibration test}
The prototype box-framing aerogels mounted in their sample aerogel panels were subjected to vibration tests using a vibration testing system at the Space Plasma Laboratory of the Institute of Space and Astronautical Science in December 2012 and April 2013. The aerogels passed the vibration tests under random vibration conditions at the levels of the acceptance test (AT) and the qualification test (QT) for launch.

To ensure the safe removal of the aerogel from the panel, the body of the panel was lined with aluminum foil (Nilaco Corporation). First, two aluminum foil sheets were interleaved and manipulated into an open box shape. The aerogel was then placed in the aluminum foil box. Finally, the aerogel was loaded in the panel by handling the aluminum foil. We confirmed that the aerogel surface had never touched the panel grid before the test. On the first attempt for the AT in December 2012, before this maneuver was implemented, the aerogel (ID = TNPP d25-1a) was accidentally partially damaged; however, we considered that the vibration test should proceed. For the QT in April 2013, we prepared a new undamaged aerogel (ID = TNPP d27-1).

\begin{figure}[t] 
\centering 
\includegraphics[width=0.48\textwidth,keepaspectratio]{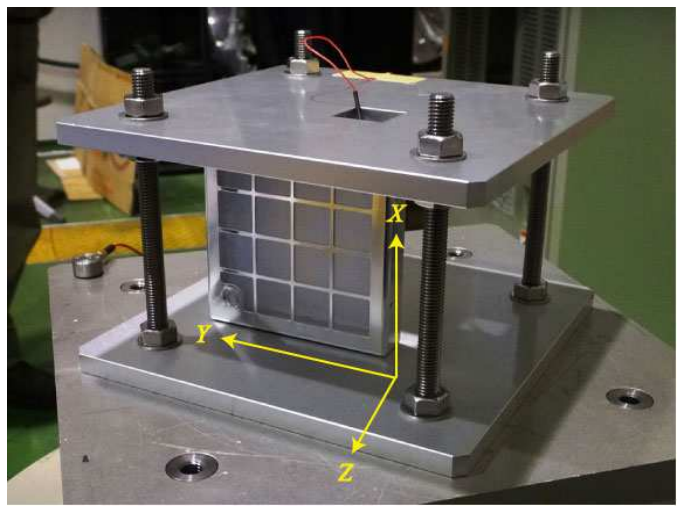}
\caption{Sample aerogel panel containing the box-framing aerogel, placed on the vibration testing station. The panel is affixed to the system stage by a special attachment. The sample is tested by vibration along the vertical axis ($X$ axis of the aerogel panel in this photograph).}
\label{fig:fig6}
\end{figure}

Vibration caused no serious damage to the aerogels. The tests were conducted under random vibration conditions at the AT and QT levels (Grms = 1.29 and 5.16 G, respectively) for launch by the H-II Transfer Vehicle or Russian Progress/Soyuz spacecraft, with the aerogel panel protected by a shock-absorbing foam of thickness 2.5 cm. In the AT, the aerogel panel alone was vibrated (without the foam) using a dedicated attachment to the testing station. In contrast, in the QT, the aerogel panel was protected by bubble wrap of thickness 2.5 cm. The vibration testing station can produce vertical vibrations only. Each test was performed along the $X$, $Y$, and $Z$ axes (see Fig. \ref{fig:fig6}), where the $Z$ axis was defined as the direction of aerogel thickness and the $X$--$Y$ plane was the horizontal plane of the aerogel. An appropriate storage container including the shock-absorbing foam to protect the aerogel panel is now under consideration.

\subsection{Pressure test}
The aerogels passed both depressurization and repressurization tests. We conducted the pressure test in a pressure-tight cylinder with valves and an oil-sealed rotary vacuum pump in May 2013. The sample aerogel panel holding the aerogel (ID = TNPP d27-1) used in the vibration test was placed in the cylinder. The cylinder was depressurized and then repressurized at the expected rate of air out-take and intake on the JEM; that is, approximately 5 kPa/s. Following the test, a measurable shrinkage was observed in the 0.01 g/cm$^3$ top layer; however, the aerogels did not fragment under the pressure changes.

\section{Conclusion}
\label{}
We are developing a silica-aerogel-based cosmic dust collector for the Tanpopo mission to be conducted on the JEM at the ISS. This paper presents the design and fabrication of the special aerogel and its holder that will be mounted on the newly developed ExHAM for the EF on the JEM.

To demonstrate the feasibility of mass production by our current technique in contamination-controlled environments (i.e., clean booth to prevent contamination from dust particles, bacterial DNA, and amino acids), we manufactured two-layer aerogels. The densities of the top and bottom layers were 0.01 and 0.03 g/cm$^3$, respectively. After resolving the layer separation problem, 106 undamaged hydrophobic aerogel products (equivalent to a yield of 84\%) were obtained by supercritical drying using carbon dioxide.

An aerogel panel of dimensions 100 mm $\times $ 100 mm $\times $ 20 mm, and a box-framing aerogel of inner and outer densities 0.01 and 0.03 g/cm$^3$, respectively, were designed and their prototypes were manufactured to interface with the ExHAM. To prevent the aerogel from escaping the panel and to ensure the safety of the ISS crew, the grid on the panel lid was designed to allow approximately 56 cm$^2$ per panel exposure area, and the maximum thickness of the aerogel was designated as 17 mm.

The ability of the box-framing aerogel mounted in the aerogel panel to withstand vibration was assessed using a vibration testing system. The aerogel suffered no serious damage under random vibration conditions at the AT and QT levels for launch. The aerogel was safely mounted (dismounted) by lining the panel body of the aerogel panel with aluminum foil. The aerogel also endured depressurization and repressurization expected in the airlock on the JEM.

\section*{Acknowledgments}
\label{}
The authors are grateful to the members of the Tanpopo working group for their assistance in this study. We are also grateful to Prof. I. Adachi of High Energy Accelerator Research Organization (KEK), the Applied Research Laboratory at KEK, Mr. M. Kubo, and Mr. T. Sato of Chiba University for their contributions to the aerogel production. The authors would like to thank Yuki Precision Co., Ltd., for their assistance in manufacturing the sample aerogel panel. This study was supported in part by the Space Plasma Laboratory at ISAS, JAXA.




\bibliographystyle{model1-num-names}



\end{document}